# Semi-Trusted Mixer Based Privacy Preserving Distributed Data Mining for Resource Constrained Devices

Md. Golam Kaosar
School of Engineering and Science
Victoria University
Melbourne, Australia
md.kaosar@live.vu.edu.au

Xun Yi, Associate Preofessor
School of Engineering and Science
Victoria University
Melbourne, Australia
xun.yi@vu.edu.au

*Abstract*— In this paper a homomorphic privacy preserving association rule mining algorithm is proposed which can be deployed in resource constrained devices (RCD). Privacy preserved exchange of counts of itemsets among distributed mining sites is a vital part in association rule mining process. Existing cryptography based privacy preserving solutions consume lot of computation due to complex mathematical equations involved. Therefore less computation involved privacy solutions are extremely necessary to deploy mining applications in RCD. In this algorithm, a semi-trusted mixer is used to unify the counts of itemsets encrypted by all mining sites without revealing individual values. The proposed algorithm is built on with a well known communication efficient association rule mining algorithm named count distribution (CD). Security proofs along with performance analysis and comparison show the well acceptability and effectiveness of the proposed algorithm. Efficient and straightforward privacy model and satisfactory performance of the protocol promote itself among one of the initiatives in deploying data mining application in RCD.

*Keywords- Resource Constrained Devices (RCD), semi-trusted mixer, association rule mining, stream cipher, privacy, data mining.*

I. INTRODUCTION

Data mining sometimes known as data or knowledge discovery is a process of analyzing data from different point of views and to deduce into useful information which can be applied in various applications including advertisement, bioinformatics, database marketing, fraud detection, e-commerce, health care, security, sports, telecommunication, web, weather forecasting, financial forecasting, etc. Association rule mining is one of the data mining techniques which helps discovering underlying correlation among different data items in a certain database. It can deduce some hidden and unpredictable knowledge which may provide high interestingness to the database owners or miners.

Rapid development of information technology, increasing use of advanced devices and development of algorithms have amplified the necessity of privacy preservation in all kind of transactions. It is more important in case of data mining since sharing of information is a primary requirement for the accomplishment of data mining process. As a matter of fact the more the privacy preservation requirement is increased, the less the accuracy the mining process can achieve. Therefore a trade-off between privacy and accuracy is determined for a particular application.

In this paper we denote Resource Constrained Device (RCD) as any kind of device having limited capability of transmission, computation, storage, battery or any other features. Examples includes but not limited to mobile phones, Personal Digital Assistants (PDAs), sensor devices, smart cards, Radio Frequency Identification (RFID) devices etc. We also interpret lightweight algorithm as a simple algorithm which requires less computation, low communication overhead and less memory and can be deployed in a RCD. Integration of communication devices of various architectures lead to global heterogeneous network which comprises of trusted, semi-trusted, untrustworthy, authorized, unauthorized, suspicious, intruders, hackers types of terminals/devices supported by fewer or no dedicated and authorized infrastructure. Sharing data for data mining purposes among such resource constrained ad-hoc environment is a big challenge itself. Preservation of privacy intensifies the problem by another fold. Therefore privacy preserving data mining in RCD envisions facilitating the mining capability to all these tiny devices which may have a major impact in the market of near future.

Data mining capability of RCD would flourish the future era of ubiquitous computing too. Owner of the device would perform mining operation on the fly. Small sensor devices would be able to optimize or extend their operations based on the dynamic circumstance instead of waiting for time consuming decision from the server. Scattered agents of a security department can take instant decision of actions about a crime or a criminal while in duty. To comprehend the necessity of lightweight privacy preserving data mining, let us consider another circumstance: there are many scattered sensor devices located in a geographical location belonging to different authorities which are serving different purposes with some common records about the environment. Now if it is required to mine data among those sensor devices to accomplish a common interest of the authorities in real time, then preserving privacy would be the first issue that must be ensured. Another motivation behind developing our proposed system could be healthcare awareness. Let us assume some community members or some university students want to know about the extent of attack of some infectious diseases such as swine flu, bird flu, AIDS etc. Each individual is very concerned about the privacy since the matter is very sensitive. They are equipped with a mobile phone or similar smart device and want to know the mining result on the fly. In such circumstances, a distributed lightweight privacy preserving data mining technique would provide a perfect solution. In addition to that; relevant people can be warned or prescribed based on all available health information including previously generated knowledge about a particular infectious diseases.

There is not much research work done for lightweight privacy preserving data mining but there is plenty of research on privacy preserving data mining. Essentially two main approaches are adapted for privacy preserving data mining solutions. First one is the randomization which is basically used for centralized data. In this approach data is perturbed using randomization function and submitted for mining. Randomization function is chosen such that the aggregated property of the data can be recognized in the miner side. In [1, 2, 3] authors have proposed such approaches. One of the major drawbacks of randomization approach is: if the precision of data mining result is increased, the privacy is not fully preserved [4].

Another one is the cryptographic approach in which the data is encrypted before it is being shared. The miner cannot decrypt individual inputs separately rather it needs to decrypt unified encrypted data together. Therefore the miner cannot associate particular information to a particular party. An example of such approach is Secure Multiparty Computation (SMC) proposed by Yao [5]. Another cryptography based privacy preservation technique is proposed by M. Kantarcioglu and C. Clifton [6] which involves enormous amount of mathematical computation and communication between data sites. This is too heavy to be implemented in a RCD. Among other privacy preserving data mining, [7] and [8] are ones which also involve vast mathematical complex equations to be solved. There are some research works on privacy issues for RCD separately too. Authors in [21] propose a technique to hide location information of a particular device for location based applications. A middleware LocServ is designed which lies in between the location-based application and the location tracking technology. A group signature based privacy for vehicles is proposed in [22], which addresses the issue of preserving privacy in exchanging secret information such as vehicle's speed, location etc.

Some research approaches address the issue of hiding sensitive information from data repository. In [23] and [24] authors basically propose some techniques to hide sensitive association rules before the data is disclosed to public. A hardware enhanced association rule mining technique is proposed in [25]. Data is needed to be fed into the hardware before the hash based association rule mining process starts. This approach may not be well realistic for RCD because it requires special purpose hardware as well as it does not handle privacy issue. A homomorphic encryption technique; Paillier encryption is used by X. Yi and Y. Zhang [9] to preserve privacy where authors propose a privacy preserving distributed association rule mining using a semi-trusted mixer. This algorithm involves lot of computation due to the use of complex mathematical equations and big prime numbers as keys in the Paillier encryption.

A heterogeneous mobile device based data collection architecture is proposed by P.P. Jayaraman [10]. Sensor devices are scattered in the environment to collect various data whereas regular mobile phones can work as bearers of the data. Detail architecture of the model is available in [10]. Authors did not consider the privacy issue during the transmission of data. If the mobile devices in the environment are intended to be utilized to work as a data bearer then privacy should be one of the major concerns. Therefore it would be difficult to be implementable in real life unless privacy is preserved. A lightweight privacy preserving algorithm similar like in this paper could provide privacy preservation as well as data mining solution for these kinds of models.

Main focus of CD [18] algorithm is to reduce communication overhead with the cost of redundant parallel computation in each data site. In addition to that this algorithm does not transmit the large itemset in the association rule mining process. Rather it communicates the counts of the itemsets only, which let it reduce communication overhead dramatically. These features make it feasible to be deployed in RCD. On the other hand semi-trusted mixer based privacy solution provided by Yi and Zhang in [9] requires lot of computation with managing big encryption key size. In this paper a more efficient semi-trusted mixer and homomorphic encryption based privacy algorithm is proposed which adopts the rule mining technique of CD to make the solution deployable in RCD.

The remainder of the paper is oriented as follows: Section 2 describes some necessary background information. Section 3 describes proposed solution which consists of privacy preserving algorithm and association rule mining algorithm for RCD. Section 4 contains security analysis and section 5 discusses the proofs and performance comparison. Finally the conclusion is presented in section 6.

## II. BACKGROUND

*Privacy*: According to The American Heritage Dictionary privacy means "The quality or condition of being secluded from the presence or view of others". In data mining if the owner of the data requires the miner to preserve privacy, then the miner gets no authority to use the data unless an acceptable and trustworthy privacy preservation technique is ensured. Different points of views define privacy in different ways. For simplicity we consider a definition which is most relevant to this work. According to J.Vaidya [1] privacy preserving data mining technique must ensure two conditions: 'any information disclosed cannot be traced back to an individual' and 'any information disclosed does not constitute an intrusion'. More technical definition of privacy can be found in [11]. This paper also provides technical definition in security analysis in section 4.

*Association Rule Mining*: Let us consider; in a distributed data mining environment collective database DB is subdivided into $DB_1$, $DB_2$, ... , $DB_N$ in wireless data sites $S_1$, $S_2$, ... ,$S_N$ respectively. I= $\{i_1, i_2, ... , i_m\}$ is the set of items where each transaction T⊆I. Typical form of an association rule is X⇒Y, where X⊆I, Y⊆I and X∩Y=ϕ. The support *s* of X⇒Y is the probability of a transaction in DB contains both X and Y. On the other hand confidence *c* of X⇒Y is the probability of a transaction containing X will contain Y too. Usually it is the interest of the data vendor to find all association rules having support and confidence greater than or equal to minimum threshold value. For another instance of an association rule AB⇒C,

$$Support_{AB \Rightarrow C} = s = \frac{\sum_{i=1}^{sites} support\_count_{ABC(i)}}{\sum_{i=1}^{sites} database\_size_{(i)}}$$

$$Support_{AB} = \frac{\sum_{i=1}^{sites} support\_count_{AB(i)}}{\sum_{i=1}^{sites} database\_size_{(i)}}$$

$$Confidence_{AB \Rightarrow C} = c = \frac{Support_{AB \Rightarrow C}}{Support_{AB}}$$

More detail on association rule mining process is available in [12, 20].

Association rule mining process consists of two major parts. First one is to find frequent large itemsets which have support and confidence values more than a threshold number of times. Second part is to construct association rules from those large itemsets. Due to the simplicity and straightforward nature of the second part, most of the association rule mining papers do not address this. Apriori algorithm is one of the leading algorithms, which determines all frequent large itemsets along with their support counts from a database efficiently. This algorithm was proposed by Agrawal in [14] which is discussed here in brief:

Let us say $L_i$ be the frequent i-itemsets. Apriori algorithm finds $L_k$ from $L_{k-1}$ in two stages: joining and pruning:

*Joining*: Generates a set of k-itemsets $C_k$, known as candidate itemsets by joining $L_{k-1}$ and other possible items in the database.

*Pruning*: Any (k−1)-itemsets cannot be a subset of a frequent k –itemset which is not frequent. Therefore it should be removed.

*Stream Cipher*: It is a symmetric key cipher where plaintext bits are combined with a pseudorandom cipher bit stream typically by an XOR operation. In stream cipher a seed is used as a key to generate continuous stream of bits. This idea can be used in generating random keys by encrypting a constant with the secret key/seed. Therefore multiple randomly generated keys can be shared among multiple entities simply by sharing a seed. In our proposed algorithm we need some randomly generated keys which can be generated by Output Feedback Mode (OFB) of Data Encryption Standard (DES) detail of which is available in [13].

*Homomorphic Encryption*: Homomorphic encryption is a special form of encryption using which one can perform a specific algebraic operation on the plaintext by performing the same or different operation on the ciphertext. Detail definition could be found in [13]. If x1 and x2 are two plaintext and E and D denotes encryption and decryption function respectively. Let us consider y1 and y2 are two ciphertexts such that: $y1=E_k(x1)$ and $y2=E_k(x2)$ where, k is the encryption key. This encryption will be considered homomorphic if the following condition is held: $y1+y2=E_k(x1+x2)$.

III. PROPOSED SOLUTION

In this paper we propose a privacy preserving secret computation protocol which is based on a homomorphic encryption technique for distributed data sites. In this section first the privacy preserving frequency mining algorithm is discussed and then the modified CD algorithm is discussed which ensures privacy in the association rule mining process.

*A. Privacy Preserving Frequency Mining*

In our proposed approach, there would be a number of participating semi honest devices or data sites (>2) which are connected to each other using heterogeneous media. There would be a semi-trusted mixer which would receive encrypted count values from sites through its private channel. It is assumed that the semi-trusted mixer would never collude with any of the data site. In practice it could be assumed that it is authorized by the government or semi-government agent. Data sites communicate to the mixer through the private channel and the mixer communicates to all sites through public channel. Necessary keys would be distributed to the sites by corresponding authorities or owner of the sites. It is also assumed that the private channel is protected by a standard secret key cryptosystem, such as DES [15] or AES [16]. Fig.1 describes the proposed model in brief.

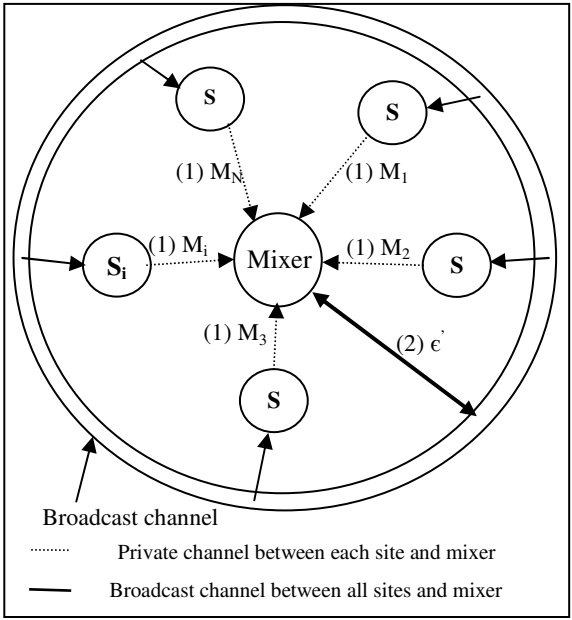

Fig.1: Privacy preserving communication and computation process between data sites and the mixer

In our proposed model we also assume no site would collude with the mixer to violate other's privacy since this would reveal privacy of itself too. In this model the privacy would be preserved if (1) the coalition of N-2 sites would not certain a revelation of privacy of any site and (2) mixer can learn nothing about the distributed database.

Let us consider; there are N resource constrained sites $S_1$, $S_2 ... S_N$ want to share the summation of a specific secret value of their own without disclosing the value itself. The secret values are $c_1, c_2 ... c_N$ respectively. $c_i^j$ denotes the value belongs to site i for $j^{th}$ iteration (in case of association rule mining it would be $j^{th}$ itemset).

*Secret parameters*: Let us assume $\rho$ is a large prime number such that, $\rho > \sum_{i=1}^{N} c_i^j$ for all j. Stream cipher seed is $\mu$. These $\rho$ and $\mu$ are shared by all the sites using any key agreement algorithm similar to one proposed in [17]. In fact there will not be any effect if $\rho$ is disclosed to the mixer. There are two more parameters $r$ and $n$ which are generated from a stream cipher in which the seed $\mu$ is used as key and any constant (may be $\rho$) as a plaintext. In each repetition the values of $r$ and $n$ will be different due to the characteristics of the stream cipher. Parameter $r$ is chosen such that $r \in Z_p^*$. If it is assumed that the length of $r$ and $n$ are $l$ bits then total number of bits in each chunk in the stream will be: $l+N.l = l(1+N)$. First $l$ bits would be the value of $r$, second $l$ bits for $n_i$ which is a random number allocated for $i^{th}$ site for communication purpose. In every iteration the value of $n_i$ would be different (similar to the value of nonce used in various cryptosystems). Thus for $j^{th}$ site $n_j$ will be allocated from bit $l+j.l$ to $l+(j+1).l$. Following figure (Fig.2) describes the allocation of values of $r$ and $n$ from the stream cipher. The length of $l$ should be chosen such that following constrained is held: $2^l - 1 = \rho$.

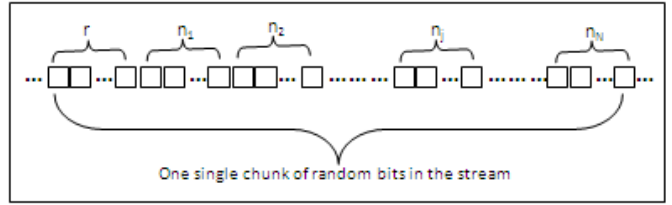

Fig.2: Random key generation from stream cipher for each iteration.

It is already mentioned that, each data sites communicate to the mixer through a private channel and the mixer communicates to all sites through public channel. Communication stages of the algorithm are depicted in the flow diagram of fig.3.

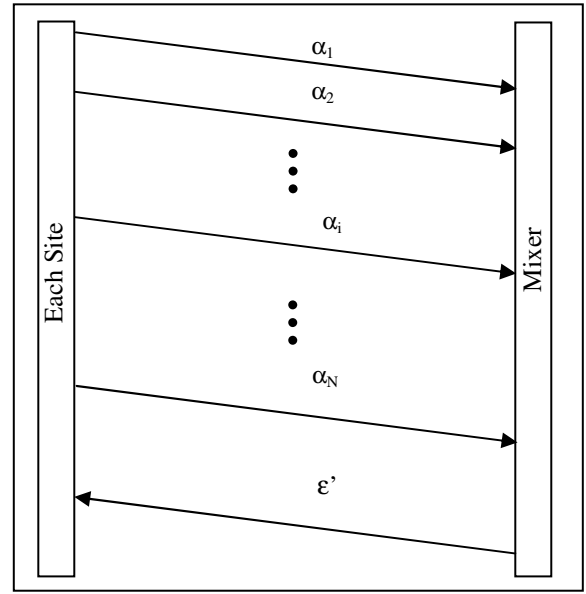

Fig.3: Flow diagram of the algorithm

*Step 1: (Encryption)*

   1.1 Each site $S_i$ computes $r_j$ following above mentioned constraints

   1.2 Encodes its count $c_i^j$ : $\alpha_i = (c_i^j . r_j + n_i) mod\ \rho$

   1.3 Then $S_i$ sends $\alpha_i$ using mixer's private channel

*Step 2: (Mixing)*

   2.1 The mixer receives $\alpha_i$ in its private channel (for all i=1 to N).

   2.2 Adds them all $\alpha$ together: $\varepsilon' = \sum_{i=1}^{N} \alpha_i$

   2.3 Broadcasts $\varepsilon'$ back to all participating sites.

*Step 3: (Decryption)*

   3.1 Each participating site $S_i$ receives $\varepsilon'$.

   3.2 $S_i$ already had computed $r_j^{-1}\ mod\ \rho$. It gets the sum of the current iteration j by computing $T_j = (\varepsilon' - R_n) . r_j^{-1}\ mod\ \rho$. Where, $R_n = \sum_{i=1}^{N} n_i$.

Thus $T_j = \sum_{i=1}^{N} c_i^j$, the sum of the count is shared among all sites without revealing individual count. An example of the algorithm is provided in the following section for more clarification.

*B. Example*

For simplicity let us consider three sites $S_1$, $S_2$ and $S_3$ want to share the sum of their count values {5, 7 and 6} without revealing their own values among themselves. Other shared and generated secret parameters are: ρ=91, r=23, $n_1$=17, $n_2$=11, $n_3$=10 and $r^{-1} = 4$ (mod 91). To minimize complexity, values of r and $n_i$ are not calculated from the stream cipher, rather their values are chosen spontaneously. Also let us assume the values of $r^{-1}$ are the same instead of different for each site. Communication between sites and the mixer is performed using private channel which is not depicted in this example too.

Exchange of count values: Each site transmits it's $\alpha_i$ to the mixer using private channel.

$\alpha_1 = (5 \times 23 + 17) mod\ 91 = 41$

$\alpha_2 = (7 \times 23 + 11) mod\ 91 = 81$

$\alpha_3 = (6 \times 23 + 10) mod\ 91 = 57$

The mixer computes

$\varepsilon' = 41 + 81 + 57 = 179$

$\varepsilon'$ is received in all sites. Sites calculate sum of counts T:

$R_n = n_1 + n_2 + n_3 = 38$

$T = (179 - 38) \times 4\ mod\ 91 = 18$

Thus T is equal to the intended sum of {5, 7 and 6}.

*C. Association Rule Mining*

Among many association rule mining algorithms we choose the one which focuses on reduction of communication cost; Parallel Mining of Association Rules [18]. In this paper authors have proposed three algorithms for the accomplishment three different objectives. Count Distribution (CD) is one of them which aims to reduce the communication cost at the cost of parallel redundant computation in each data site. In this subsection we would integrate our proposed privacy preserving communication technique with CD algorithm which would be suitable for RCD in terms of computation and communication.

Since frequent large itemset computation is considered as the major task in association rule mining algorithms, we focus our effort for the accomplishment of the same task as it is the case in many other papers. Following are the notations, major stages and actions performed in each data site in every cycle:

Let, $S_i$: Data site (site) of index i. N: Number of sites. $DB_i$: Database (collection of transactions) in $S_i$. $L_k$: Set of frequent k-itemset. $C_k$: Set of candidate k-itemset and $DB = \sum_{i=1}^{N} DB_i$.

(1) *Candidate set generation*: Each site $S_i$ generates a complete candidate set $C_k$ from $L_{k-1}$ which is computed in the previous iteration using Apriori algorithm (discussed in section 2).

(2) *Count computation*: $S_i$ passes over all the transactions in $DB_i$ to compute the count for all items in $L_{k-1}$.

(3) *Transmission of counts*: Counts of $C_k$ is sent to the mixer using privacy preserving communication techniques discussed in subsection 3.1. Communication between the data sites and the mixer is performed through the private channel. The value of j in the algorithm (subsection 3.1) maps to the itemsets sequence number.

(4) *Mixer functions*: Mixer adds all the encrypted counts received from all the sites and broadcasts the result back to all sites.

(5) *Result decryption*: Each data site decrypts the result received from the mixer as it is stated in section 3.1 to get sum of the counts.

(6) *Termination*: Since all sites perform identical operation, all of them terminate at the same iteration and end up with generation of large itemset.

IV. SECURITY ANALYSIS

In this section we demonstrate that our proposed protocol preserves privacy during the transmission of counts of itemsets in association rule mining process. With the basis of privacy requirement and security definition provided in [9, 19], following formulation can be addressed.

Let us assume N≥3, since privacy preservation is impossible for less than three parties. VIEW($S_i$, N) implies view of the party $S_i$ where total number of participants is N. Similarly VIEW(M,N) implies the view of the mixer. Therefore by definition VIEW(M,0), VIEW($S_i$,0), VIEW($S_i$,1) and VIEW($S_i$,2) all equal to Φ. If X and Y are two random variables then,

X$\approx$<sub>poly</sub>Y = (the probability of distinguishing X and Y) $\leq \frac{1}{2} + \frac{1}{Q(l)}$ for all polynomials Q(*l*) [9]. N parties want to find the sum of their counts of itemset $c_1, c_2 \ldots c_N$. The privacy will be preserved if following conditions are satisfied [9].

(a) Two random variables $A_{N,j} = (VIEW(M,N), \sum_{i=1}^{j} c_i)$ and $B_{N,j} = (VIEW(M,N), R)$ are polynomially indistinguishable ($A_{N,j}\approx_{poly}B_{N,j}$) for 1≤j≤N and 0≤R<ρ.

(b) Two random variables $C_{N,j} = (\cup_{i=1}^{j} VIEW(S_i, N), c_{j+1})$ and $D_{N,j} = (\cup_{i=1}^{j} VIEW(S_i, N), R)$ are polynomially indistinguishable ($C_{N,j}\approx_{poly}D_{N,j}$) for n ≥ 3, 1 ≤ j ≤ n-2 and 0≤R<ρ.

Since all users have identical values of $(VIEW(M,N), \sum_{i=1}^{j} c_i)$, $(VIEW(M,N), \sum_{i=2}^{j} c_i)$ … $(VIEW(M,N), \sum_{i=N-j+1}^{N} c_i)$, they are the same.

*Theorem 1*: The proposed protocol preserves privacy based on the above mentioned privacy definition.

*Proof*: (a) When N=1, then j=1 and $A_{1,1} = (VIEW(M,1), c_1) = (\alpha, c_1)$.

Since the mixer does not know the secret parameters ($\rho$, $\mu$) it cannot decrypt $\alpha$. Therefore
$A_{1,1} = (\alpha, c_1) \approx_{poly} (\alpha, r) = (VIEW(M,1), r) = B_{1,1}$.

When N>1 and 1≤j≤N

$$A_{N,j} = \left(VIEW(M,N), \sum_{i=1}^{j} c_i\right)$$

$$= \left(VIEW(M, N-j), view(M,j), \sum_{i=1}^{j} c_i\right)$$

$$= \left(VIEW(M, N-j), E\left(\sum_{i=1}^{j} c_i\right), \sum_{i=1}^{j} c_i\right)$$

$\approx_{poly} (VIEW(M, N-j), E(\sum_{i=1}^{j} c_i), R)$ [Since $A_{1,1} \approx_{poly} B_{1,1}$]
$= (VIEW(M, N-j), R) = B_{N,j}$

(b) When n=3, j=1. Therefore

$$C_{3,1} = (VIEW(S_1, 3), c_2) = \left(\left(c_1, \sum_{i=1}^{3} c_i\right), c_2\right)$$

With given $c_1$ and $\sum_{i=1}^{3} c_i$, party $S_1$ cannot be certain about $c_2$. Therefore,

$C_{3,1} \approx_{poly} (c_1, \sum_{i=1}^{3} c_i, R) = (VIEW(S_1, 3), R) = D_{3,1}$

When N>3 and 1≤ j ≤ n-2,

$C_{N,j} = (\cup_{i=1}^{j} VIEW(S_i, N), c_{m+1})$

$$= \left(\left(\sum_{i=1}^{j} c_i, \sum_{i=1}^{N} c_i\right), c_{m+1}\right)$$

Let us assume $c'_1 = \sum_{i=1}^{j} c_i$, $c'_2 = c_{j+1}$, $c'_3 = \sum_{i=1}^{N} c_i - \sum_{i=1}^{2} c'_i$

Since $C_{3,1} \approx_{ploy} D_{3,1}$

$C_{N,j} = \left(c'_1, \sum_{i=1}^{3} c'_i, c'_2\right) \approx_{poly} \left(c'_1, \sum_{i=1}^{3} c'_i, R\right)$

$= \left(\sum_{i=1}^{j} c_i, \sum_{i=1}^{N} c_i, R\right) = (\cup_{i=1}^{m} VIEW(S_i, N), R) = D_{N,j}$

Therefore the privacy is preserved for the proposed protocol.

*Theorem 2*: The protocol does not reveal support count of one participant to either the mixer or to other participants.

*Proof*: In step 1.2 of the algorithm, each site $S_i$ encrypts the secret value using private keys which are only known to sites. Before the ciphertext is transmitted to the private channel of the mixer, it is farther encrypted using the public key of the mixer in step 1.3. None has the private key of the mixer except the mixer itself; therefore no eavesdropper can get access to the ciphertext. On the other hand the mixer only can decrypt the outer encryption of the double encrypted ciphertext. It cannot decrypt or read the secret value of $S_i$. Mixer only adds all the ciphertexts together and broadcasts the result to all sites in step 2. Now the sum is known to all parties. They all can decrypt it which is a summation of their secret values. Therefore none can reveal or relate any secret value associated to any site.

*Theorem 3*: Security against the mixer and any other individual participant or outsider

*Proof*: Unlike any other kind of regular security protocols our proposed protocol has neither a straight forward sender nor a receiver. Rather it involves encryption of different contents with different keys by multiple senders, a mixer and multiple receivers together in a complete single communication. The senders send in the first step and receive in the third step. Moreover each transaction in this protocol is consists of multiple communication attempts, which make the protocol different and more secure compared to other protocols. Let us study the vulnerability in following cases:

*Replay attack*: If an eavesdropper gets all the communications between all sites and the mixer, he cannot learn anything significant about the secret value of an individual party. Because in every communication the value of $n_j$ chosen randomly in step 1.2 of the algorithm, which would raise the high degree of unpredictability of the data in the channel.

*Brute force attack*: Again due to the frequent and random change of value of $n_j$ in each communication, brute force attack is unrealistic.

V. PERFORMANCE ANALYSIS

Yi-Zhang's [9] privacy preserving association rule mining algorithm uses semi-trusted mixer which is similar to our proposed model. We compare the performance of the proposed protocol with Yi-Zhang protocol. To measure and compare the performance between these two protocols, let us assume following parameters:

H= Average number of items in the large k-itemset.
L= Size of each entry in the list of large itemsets to store index and count in Bytes.
N= Number of data sites.
K= Average size of each item in Byte (number of characters as for example).
$\varphi$=Encryption ratio ($\frac{Cipher-text\ size}{Plain-text\ size}$) in step 1.2 in the proposed algorithm.
| $\alpha_i$|=Size of $\alpha_i$ in step 1.2 in the proposed algorithm.
|ϵ'|= Size of ϵ' in step 2.3 in the proposed algorithm.

*Proposed algorithm*: Communication payload in each iteration is

N*| $\alpha_i$|*H+|ϵ'|*H*N = N*$\varphi$*L*H+$\varphi$*H*N = $\varphi$HN(1+L)

In case of $\varphi$=1, Communication overhead= HN(1+L).

*Yi-Zhang algorithm [9]*: Let us assume same encryption ratio (that is same φ and β') in both level of encryptions. Communication payload in each iteration is:

N*(Cipher-text sent by each site) + Data broadcasted by the mixer = N*φ*H*K+N* φ*H*K=2φNHK.

If φ=1, Communication overhead= 2NHK.

For farther comparison let us assume value of L=2 (two bytes to store two values: index and count) and K=3 (on an average). Therefore communication payload in our proposed algorithm and Yi-Zhang's algorithm are 3NH and 6NH bytes respectively. Therefore the proposed algorithm generates as much as half communication payload of the Yi-Zhang algorithm.

Let us now compare the number of instructions necessary in encrypting and decrypting a message m. We compare only the homomorphic encryption involved in both Yi-Zhang and the proposed protocol. Basic encryption and decryption equations of Yi-Zhang protocol are:

Encryption: $c = g^m r^N (mod\ N^2)$ and

Decryption: $m = \frac{(c^\lambda (mod\ N^2)-1)/N}{(g^\lambda (mod\ N^2)-1)/N} (mod\ N)$

Where, m: the message, c: the ciphertext, N: pq (p and q are large prime numbers), g: public key, r: a random number.

Therefore number of operations involved for encryption and decryption are:

Exponential operations: 1+1+1+1=4

Basic operations: 1+1+1+1+1+1+1+1+1+1+1+1+1=13

In case of the proposed protocol, basic encryption and decryption equations are (as stated in section 3.1):

Encryption: $c = (m.r + n) mod\ \rho$ and

Decryption: $m = (c - r).r\ mod\ \rho$

Where, r and n: random numbers, ρ: prime number > sum of counts of items. For the sake of measuring the operations count, we treat $\sum r$ and r as the same.

Therefore number of basic instructions involved in encryption and decryption are:

Exponential operations: 0

Basic operations: 1+1+1+1=4

Finally let us compare the size of the keys used in both the protocols:

In the proposed protocol the key is considered as the seed μ of the stream cipher. The size of μ can be considered as a typical one: 80 bits.

Yi-Zhang protocol: It is mentioned in [9] that for security concern the value of N should be such that $log_2^N \approx 1024$. Therefore size of N is 1024 bits.

All the performance comparisons between Yi-Zhang and the proposed protocol are summarized in table 1.

| Measure | Our Proposed Protocol | Yi-Zhang Protocol |
|---|---|---|
| Communication overhead (each iteration) | 3NH | 6NH |
| Number of exponential operations | 0 | 4 |
| Key size | 80 | 1024 |

Table 1: Performance comparison between Yi-Zhang and the proposed protocol.

Though there is no use of exponent operations in the proposed algorithm, it involves some other cryptographic operations which would be efficient enough due to small key size. Therefore the performance comparison shows that the proposed algorithm is more efficient and straightforward, which make it suitable for RCD.

## VI. CONCLUSION

Rapid development and increasing popularity of ubiquitous computing and RCD in the environment demands the deployment of varieties of lightweight applications. A lightweight algorithm which would lead one step ahead to deploy data mining applications in RCD is proposed in this paper. All security protocols involve detail consideration of various security threats. But our proposed model can avoid many security threats such as replay attack, brute force attack etc, due to the nature of the protocol itself. This is so because in this protocol a single communication is not consists of simply between a sender and a receiver rather it involves multiple senders, receivers and the mixer all together. All the secret parameters and keys in our proposed homomorphic encryption technique are very small in size; therefore less computation is involved in the encryption and decryption process. This feature makes the proposed algorithm more suitable for RCD. Performance analysis and proofs of privacy and security also imply the strength and appropriateness of the algorithm. Therefore this effort should be considered as one of the effective initiative towards the deployment of data mining in ubiquitous computing environment.

## REFERENCES

[1] R. Agrawal, R. Srikant, "Privacy-preserving data mining", Proceedings of ACM SIGMOD International Conference on Management of Data, 2000, pp. 439–450.

[2] A. Evfimievski, "Randomization in privacy preserving data mining", ACM SIGKDD Explorations Newsletter, Volume 4 Issue 2, December 2002, pp. 43 – 48

[3] A. Evfimievski, S. Ramakrishnan, R. Agrawal, J. Gehrke, "Privacy-preserving mining of association rules", 8th ACM SIGKDD International Conference on Knowledge Discovery and Data Mining, ACM Press, 2002, pp. 217–228.

[4] H. Kargupta, S. Datta, Q. Wang, K. Sivakumar, "On the privacy preserving properties of random data perturbation techniques" 3rd Int'l Conference on Data Mining, 2003, pp. 99–106.

[5] A.C. Yao, "How to generate and exchange secrets", 27th IEEE Symposium on Foundations of Computer Science, 1986, pp. 162–167.

[6] M. Kantarcioglu, C. Clifton,"Privacy-preserving distributed mining of association rules on horizontally partitioned data", Knowledge and Data Engineering IEEE Transaction Volume 16, Issue 9, Sep. 2004 pp. 1026-1037.

[7] Y. Lindell, B. Pinkas, "Privacy Preserving Data Mining", Journal of Cryptology, Volume 15 - Number 3, 2002.

[8] Z. Yang, S. Zhong, R. N. Wright, "Privacy-Preserving Classification of Customer Data without Loss of Accuracy", Proceedings of the Fifth SIAM International Conference on Data Mining, Newport Beach, CA, April 21-23, 2005.

[9] X. Yi, Y. Zhang, "Privacy-preserving distributed association rule mining via semi-trusted mixer", Data and Knowledge Engineering, vol. 63, no. 2, 2007.

[10] P.P. Jayaraman, A. Zaslavsky, J. Delsing, "Sensor Data Collection Using Heterogeneous Mobile Devices", Pervasive Services, IEEE International Conference, Istambul, 15-20 July 2007 pp. 161-164.

[11] J. Vaidya, C. Clifton, M. Zhu, "Privacy Preserving Data Mining", Springer 2006, ISBN-13: 978-0-387-25886-8.

[12] J. Han, M. Kamber, "Data Mining Concepts and Techniques", Second Edition, Elsevier Inc. 2006. ISBN: 13:978-1-55860-901-3.

[13] J. Katz, Y. Lindell, "Introduction to Modern Cryptography", Taylor & Francis Group, LLC, 2008. ISBN: 13: 978-1-58488-551-1.

[14] R. Agrawal and R. Srikant, "Fast algorithms for mining association rules," in Proceedings of the 20th International Conference on Very Large Data Bases. Santiago, Chile: VLDB, Sept. 12-15 1994, pp. 487–499.

[15] NBS FIPS PUB 46, Data Encryption Standard (National Bureau of Standards, US Department of Commerce, 1977).

[16] FIPS PUB 197, Advanced Encryption Standard (Federal Information Processing Standards Publications, US Department of Commerce/N.I.S.T., National Technical Information Service, 2001).

[17] S. B. Wilson, A. Menezes, "Authenticated Diffie-Hellman Key Agreement Protocols", Lecture Notes in Computer Science, Springer-Verlag Berlin Heidelberg, ISBN- 978-3-540-65894-8, January 1999, pp. 339-361.

[18] R. Agrawal and J.C. Shafer, "Parallel Mining of Association Rules", Knowledge and Data Engineering, IEEE Transactions on Volume 8, Issue 6, Dec. 1996 pp. 962-969.

[19] W.G. TZeng, "A secure fault-tolerant conference key agreement protocol", IEEE Transactions on Computers vol. 51 issue 4, April 2002, pp. 373 – 379.

[20] P-N. Tan, M. Steinbach, V. Kumar, "Introduction to Data Mining", 1st Edition, ISBN/ISSN: 0321321367, 2006.

[21] G. Myles, A. Friday, N. Davies, "Preserving privacy in environments with location-based applications", Pervasive Computing IEEE, vol. 2, issue 1, Jan-Mar 2003, pp 56-64.

[22] J. Guo, J. P. Baugh, S. Wang, "A Group Signature Based Secure and Privacy-Preserving Vehicular Communication Framework" 2007 Mobile Networking for Vehicular Environments, May 2007, pp 103-108.

[23] V.S. Verykios, A.K. Elmagarmid, E. Bertino,Y. Saygin, E. Dasseni, "Association rule hiding", Knowledge and Data Engineering, IEEE Transactions, Volume 16, Issue 4, April 2004, pp. 434 – 447.

[24] B.J. Ramaiah, A. Reddy, M.K. Kumari," Parallel Privacy Preserving Association rule mining on pc Clusters", 2009 IEEE International Advance Computing Conference (IACC 2009), March 2009, pp. 1538 – 1542.

[25] Y.H. Wen, J.W. Huang, M.S. Chen, "Hardware-Enhanced Association Rule Mining with Hashing and Pipelining", IEEE Transactions on Knowledge and Data Engineering, vol. 20, no. 6, June 2008.

AUTHORS PROFILE

**Md. Golam Kaosar** is a PhD student at the School of Engineering and Science, Victoria University, Australia. Before he starts his PhD, he used to work as an engineer at Research Institute (RI) in King Fahd University of Petroleum and Minerals (KFUPM), Saudi Arabia. Before that he got his MS in Computer Engineering and BSC in Computer Science and Engineering from KFUPM, and Bangladesh University of Engineering and Technology (BUET), Bangladesh at the years 2006 and 2001 respectively.

As a young researcher, he has a good research background. He has published number of conference papers including IEEE and some journals. His area of research includes but not limited to Privacy Preserving Data Mining, Ubiquitous Computing, Security and Cryptography, Ad-hoc sensor network, Mobile and Wireless Network, Network Protocol, etc.